\documentclass[prd,aps,twocolumn,showpacs,amsmath,amssymb,floatfix,nofootinbib]{revtex4}
\usepackage{amsfonts}
\usepackage[colorlinks=true, citecolor=blue, linkcolor=blue, urlcolor=blue]{hyperref}
\usepackage{color}
\usepackage{multirow}
\usepackage{graphicx,amsfonts,multirow}
\usepackage{amsmath}
\usepackage{ulem}

\newcommand\grap[2]{\includegraphics[width=#1\textwidth]{#2}}
\newcommand\dz[2]{\delta Z_#1^{\rm #2} }
\newcommand\epx[1]{\epsilon_{\rm #1}}
\newcommand\lnn[1]{\ln \frac{4\pi \mu_r^2}{#1^2}}

\newcommand{\gE}{\gamma_E}
\newcommand{\as}{\alpha_s}
\newcommand{\bo}{\beta_0}
\newcommand{\CF}{C_F}
\newcommand{\CA}{C_A}
\newcommand{\TF}{T_F}
\newcommand{\nn}{\nonumber}
\newcommand{\hl}{\hline}

\newcommand{\G}{\Gamma}
\newcommand\g[1]{\gamma^#1}

\newcommand{\beq}{\begin{eqnarray}}
\newcommand{\eeq}{\end{eqnarray}}

\newcommand{\bfig}{\begin{figure}}
\newcommand{\efig}{\end{figure}}

\newcommand{\bfigs}{\begin{figure*}}
\newcommand{\efigs}{\end{figure*}}

\newcommand{\btab}{\begin{table}}
\newcommand{\etab}{\end{table}}

\newcommand{\btabu}{\begin{tabular}}
\newcommand{\etabu}{\end{tabular}}

\newcommand{\btabs}{\begin{table*}}
\newcommand{\etabs}{\end{table*}}

\usepackage{epstopdf,multirow}
\usepackage{color,xcolor}
\definecolor{rred}{rgb}{0.7,0,0}
\allowdisplaybreaks
\date{\today}

\begin{document}

\title{Next-to-leading-order QCD correction to the exclusive double charmonium production via $Z$ decays}
\author{Xuan Luo}
\email{cnluoxuan@hotmail.com}
\author{Hai-Bing Fu}
\email{fuhb@gzmu.edu.cn}
\author{Hai-Jiang Tian}
\author{Cong Li}
\address{Department of Physics, Guizhou Minzu University, Guiyang 550025, People's Republic of China}

\begin{abstract}
In this paper, we preformed a further research on the exclusive productions of double charmonium via $Z$-boson decay by using nonrelativistic QCD factorizations approach, where the single-photon fragmentation topologies of the QED diagrams, the interference terms between the QCD and full QED diagrams, the next-to-leading-order calculations of the interference terms are preformed. For the production of $J/\psi+J/\psi$ in $Z$-boson decay, the interference terms show a significantly phenomenological effect due to the addition of the newly calculated NLO QCD corrections. After adding together all contributions, the branching fraction $\mathcal{B}_{Z\to J/\psi J/\psi}$ still undershoots the CMS collaboration data obviously. In addition, we simultaneously complete the next-to-leading-order calculations for $Z\to J/\psi+\eta_c (\chi_{cJ})$ with $J=(0,1,2)$. The calculated results show that the newly-calculated complete QED and cross terms will have obvious effective on the total decay widths.
\end{abstract}

\date{\today}

\maketitle

\section{Introduction}
In heavy charmonium physics, the investigation for double charmonium states catches the attention of physicist, due to the theoretical predictions of its production process does not include any unknown non-perturbative parameters. Over the past decades, the research on double charmonium at various high-energy colliders has been made considerable progress both experimentally and theoretically~\cite{Braaten:2002fi, Bodwin:2002kk, Gong:2008ce,Likhoded:2016zmk, He:2007te, Zhang:2005cha, Borschensky:2016nkv, Baranov:2015cle, Qiao:2014pfa, Lansberg:2014swa, Li:2013otv, Qiao:2002rh, BrennerMariotto:2018eef, Sun:2014gca, NA3:1982qlq, BaBar:2005nic, Belle:2002tfa, Belle:2004abn, D0:2014vql, LHCb:2011kri, CMS:2014cmt, Feng:2019zmt, Sun:2018rgx, Jiang:2018wmv}. On the experimental side, the $J/\psi+\eta_c$ pair and $J/\psi+\chi_{c0}$ pair have been collected in $e^+e^-$ collision\cite{BaBar:2005nic, Belle:2002tfa}, and the LHCb, CMS and D0 collaborations reported the observations of the hadroproduction of $J/\psi+J/\psi$ pair~\cite{D0:2014vql,LHCb:2011kri,CMS:2014cmt}. Theoretically, the corrections is reaches next-to-next-to-leading-order accuracy for the process $e^+e^-\to J/\psi+\eta_c$~\cite{Feng:2019zmt}, while the predicts of $J/\psi+\chi_{cJ}$ with $J=(0,1,2)$ via $e^+e^-$ annihilation is up to next-to-leading-order (NLO) QCD corrections~\cite{Jiang:2018wmv}. More double charmonium pair in $e^+e^-$ annihilation production at LO level has been discussed in Ref.~\cite{Chen:2013mjb}.

To better understanding the mechanism of the double charemonium production, in addition to direct production, double charmonium indirect production is also interest, which may inform us not only the character of double charmonium, but also the properties of its parent particles. For instance the direct production of double charmonium via Higgs boson and $Z$-boson decay. Among of them, heavy-quarkonium production via $Z$-boson decay is a particularly example that can offer a reference for recognition the different models, the details can be found in the literatures~\cite{Cheung:1995ka, Cho:1995vv, Baek:1996np, Bodwin:1994jh}.

The CMS collaboration at LHC has released the upper limits on the branching fractions for the $Z\to J/\psi+J/\psi$ (i.e, $\mathcal{B}_{Z\to J/\psi J/\psi}<2.2\times 10^{-6}$) in 2019~\cite{CMS:2019wch}, and the branching fractions has been update to be $\mathcal{B}_{Z\to J/\psi J/\psi}<1.1\times 10^{-6}$ by CMS Collaboration in 2022~\cite{CMS:2022fsq}. The LO predictions obtained by calculating the pure QCD process are only at the $10^{-12}$ order~\cite{Likhoded:2017jmx}, which are significantly undershoot the CMS measurements. Recently, in Ref.~\cite{Gao:2022mwa}, the authors raised the magnitude of the upper limits on the branching fractions for the $Z\to J/\psi + J/\psi$ to $10^{-10}$ by introducing a single-photon fragmentation (SPF) mechanism at LO level, which is certainly a huge breakthrough. However, the discrepancies between theory and CMS collaboration data still obvious.

Reviewing the exclusive $J/\psi+\eta_c(\chi_{cJ})$ productions via $e^-e^+$ annihilation~\cite{Zhang:2005cha,Zhang:2008gp}, the NLO QCD correction to the pure QCD process significantly reduced the gap between theoretical predictions and experimental results. Moreover, based on fact that suppressed by $\alpha^2/\as^2$, the QED process had been overlooked. But, as the pointed out in Ref.~\cite{Bodwin:2002fk}, the QED process can also provide obvious contributions at LO level, due to the large kinematic elevate caused through the SPF topology of the QED diagram. In Refs.~\cite{Sun:2018rgx,Jiang:2018wmv}, the cross terms between the complete QED and QCD LO-level diagrams and the NLO QCD corrections to these cross terms can significantly effect of the predicts. In conclusion, its very essential to consider the NLO QCD correction of the cross terms in the process of double charmonium production after introduction of the QED diagram.

\bfigs[t]
\centering
\grap{0.76}{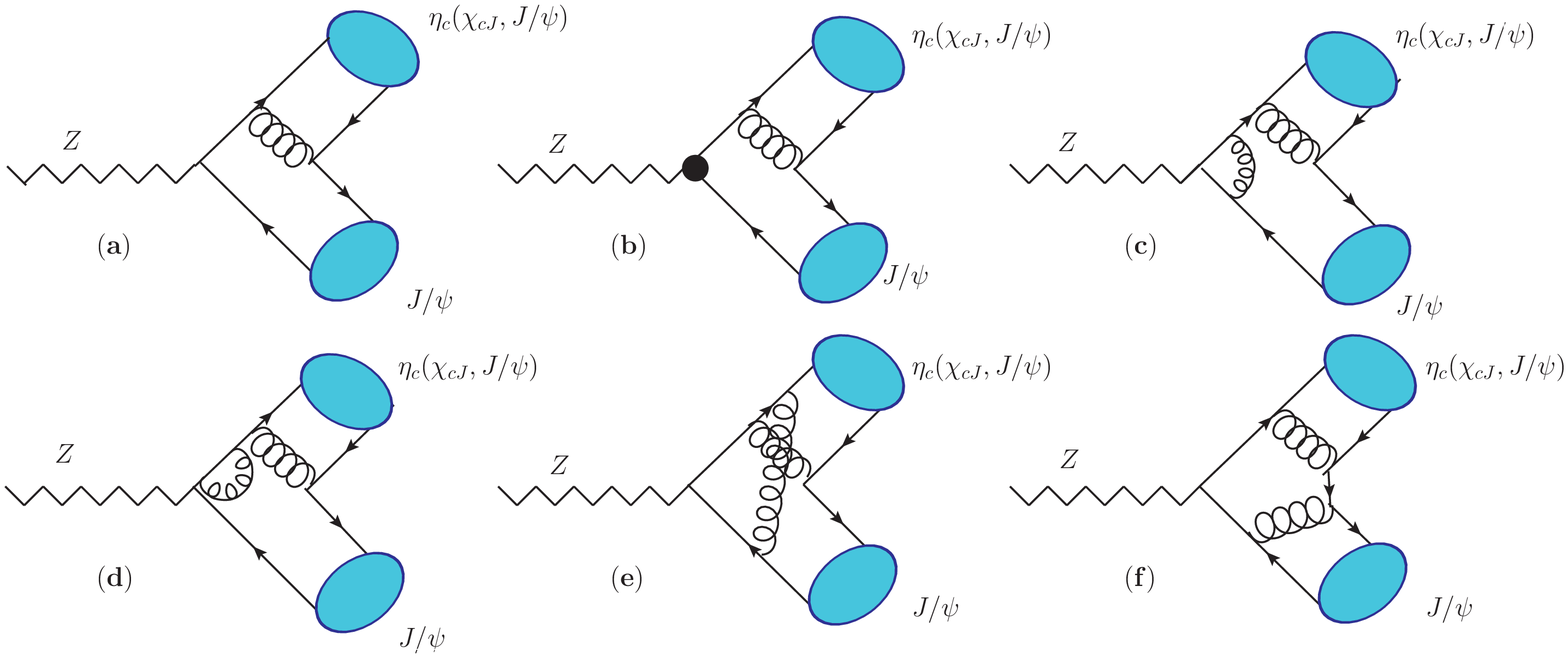}
\caption{(Color online) The first group typical Feynman diagrams for $Z\to J/\psi+\eta_c(J/\psi,\chi_{cJ})$ with $J=(0,1,2)$. in which (a) represents the QCD diagram at leading-order level, (b)-(f) denotes next-to-leading-order QCD corrections of diagram (a), and (b) is counter term diagram.}\label{tu1}
\efigs
\bfigs[t]
\centering
\grap{0.99}{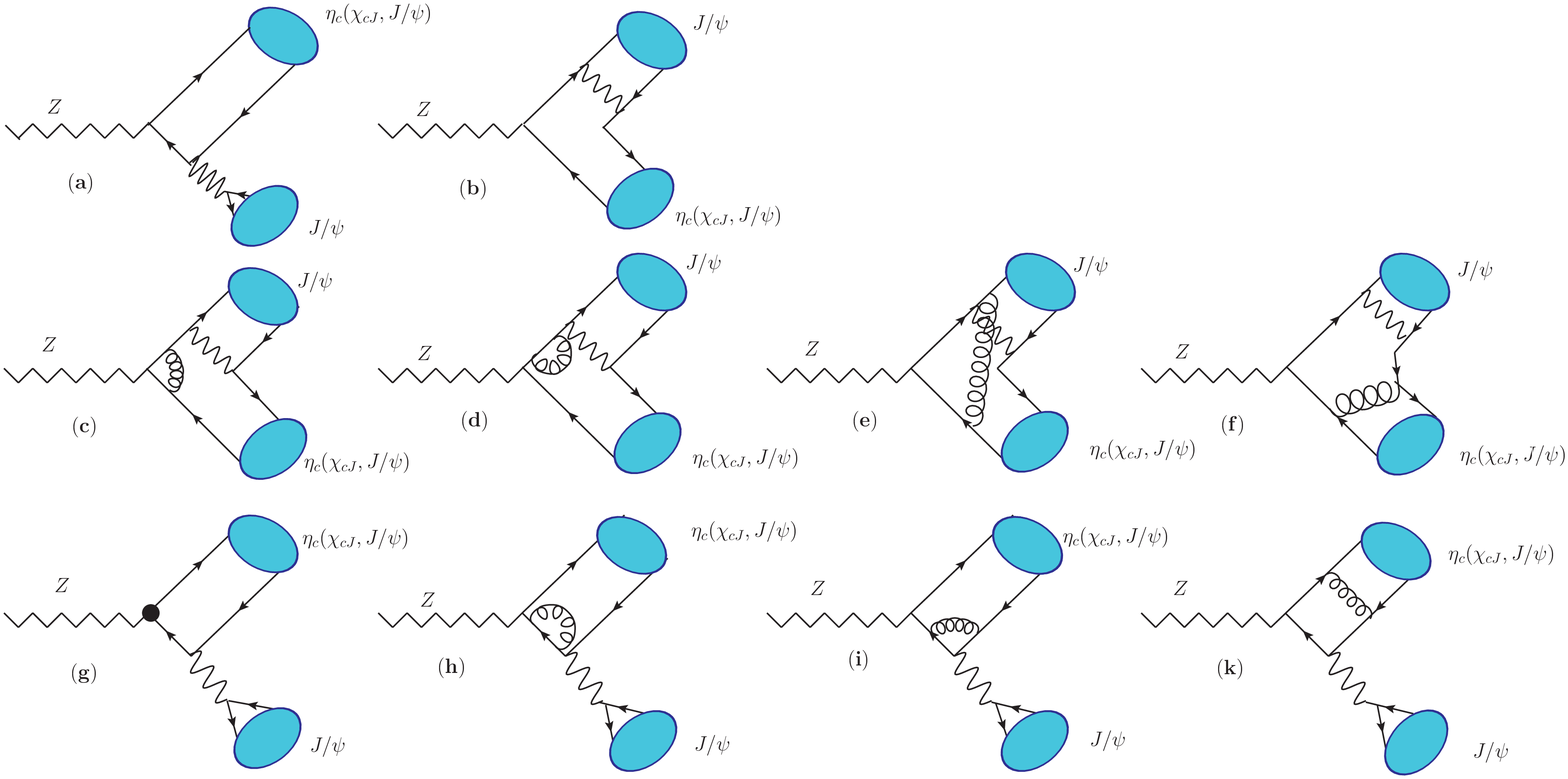}
\caption{(Color online) The second group typical Feynman diagrams for $Z\to J/\psi+\eta_c(J/\psi,\chi_{cJ})$ with $J=(0,1,2)$. (a) and (b) are the SFP diagram and usual QED diagram at leading-order level, respectively. (c)-(f) denotes NLO QCD corrections of (a) and (g)-(k) denotes NLO QCD corrections of (a). The (g) is counter term diagram.}\label{tu2}
\efigs

Due to there exist the SPF topology in $Z\to J/\psi+J/\psi$ process, the NLO QCD correction of these cross terms between QCD and QED diagrams may further narrow the gap between theoretical predictions and experiments and is worth investigating. In addition to $J/\psi+J/\psi$, $J/\psi+\eta_c(\chi_{cJ})$ with opposite charge conjugation can also be obtained via $Z$-boson decay, and is deserving investigation as well. In order to achieve this goal, we will performed a research on the indirect production of $J/\psi+J/\psi(\eta_c,\chi_{cJ})$ via $Z$-boson decay at NLO level by introducing the SPF topologies of the QED diagrams for the first time, to further alleviate the tension between theoretical results and CMS Collaboration data $\mathcal{B}_{Z\to J/\psi J/\psi}$.

The remaining components of the paper are organized as follows: In Sec.~\ref{section:2}, we demonstrate the calculation technology in detail. Numerical analysis and discussions are given in Sec.~\ref{section:3}. In Sec.~\ref{section:4}, the summary and conclusion are listed.

\section{General formalism}\label{section:2}

Based on the nonrelativistic QCD (NRQCD) approach~\cite{Bodwin:1996tg,Petrelli:1997ge}, the differential decay width for the indirect production of double charmonium in $Z$-boson decay can be explicitly written as
\beq
d\G=\sum_{n}d\hat\G(Z \to H_1+H_2)\langle {\cal O}^H(n) \rangle.
\label{eq:1}
\eeq
In which the $H_i$ with $i=(1,2)$ stand for the charmonium. $\langle {\cal O}^H(n)\rangle $ is the non-perturbative long-distance matrix element, which can be obtained by approximatively related to the origin value of the Schr\"{o}dinger wave function or its derivative, $\langle {{\mathcal O^H}(n)}\rangle =|\Psi_{c\bar c}(0)|$ for $S$-wave, or $\langle {\cal O}^H(n)\rangle =| \Psi^\prime_{c\bar c}(0) |$ for $P$-wave. ${| {{\Psi_{c\bar c}}(0)} |} $ and ${| {{\Psi^{\prime}_{c\bar c}}(0)} |}$ for hadrons are fitted from the experiment data or some theoretical approaches, e.g. the potential model \cite{Bagan:1994dy}, lattice QCD (LQCD) \cite{Bodwin:1996tg}, or QCD sum rules \cite{Kiselev:1999sc}.
The decay width $d\hat\G(Z \to H_1 + H_2)$ can be written as
\beq
d\hat\G(Z \to H_1+H_2) = \frac13 \frac1{2m_Z} \sum |M[n]|^2 d\Phi_3. \label{eq:2}
\eeq
After considering the complete QED process, the $|M[n]|^2$ have following expression,
\beq
&&\hspace{-0.7cm}| (M_{a^{1/2}a_s} + M_{a^{1/2}a_s^2}) + (M_{a^{3/2}} + M_{a^{3/2}a_s})|^2
\nn\\
&&\quad= | M_{a^{1/2}a_s}|^2 + |M_{a^{3/2}}|^2 + 2{\rm Re} (M_{a^{1/2}a_s}M_{a^{3/2}}^*)
\nn\\
&&\quad+2{\rm Re} (M_{a^{1/2}a_s^2}M_{a^{1/2}a_s}^*) + 2{\rm Re} (M_{a^{1/2}a_s^2}M_{a^{3/2}}^*)
\nn\\
&&\quad+ 2{\rm Re} (M_{a^{1/2}a_s}^* M_{a^{3/2}a_s}) + \cdots~.
\label{eq:1}
\eeq
According to the Feynman rule, one can get the 86 QCD diagrams for the processes $Z\to J/\psi + H$ with $H = (\eta_c, J/\psi, \chi_{cJ})$. In which, the numbers of Born, one-loop and counter-terms diagrams are 4, 62 and 20, respectively. Then, there have 72 QED diagrams for the $Z\to J/\psi +\eta_c(\chi_{cJ})$, including 6 Born, 42 one-loop and 24 counter-terms diagrams, respectively. The numbers of QED diagrams for the $Z\to J/\psi +J/\psi$ is 92, including 8 Born, 52 one-loop and 32 counter-terms diagrams, respectively. Partial Feynman diagrams of QCD process at LO level and its QCD correction at NLO level are presented in Fig.~\ref{tu1}. The Feynman diagrams of QED process at LO level and its QCD correction at NLO level are listed in Fig.~\ref{tu2}. Omit the higher-order terms in $\as$, the differential decay width can be divided into the following five components
\beq
d\G = d\G_1^0+d\G_2^0 + d\G_2^1 + d\G_3^0 + d\G_3^1, \label{eq:1}
\eeq
in which, each terms can be written as
\beq
&&\hspace{-0.5cm}d\G_1^0 \propto | M{_{a^{3/2}}} |^2,\\
&&\hspace{-0.5cm}d\G_2^0 \propto | M{_{a^{1/2}{a_s}}} |^2,\\
&&\hspace{-0.5cm}d\G_2^1 \propto 2{\rm Re} (M{_{a^{1/2}a_s^2}} M_{a^{1/2} a_s}^*),\\
&&\hspace{-0.5cm}d\G_3^0 \propto 2{\rm Re} (M{_{a^{1/2}a_s}}M_{a^{3/2}}^*),\\
&&\hspace{-0.5cm}d\G_3^1 \propto 2{\rm Re} (M{_{a^{1/2}a_s^2}}M_{a^{3/2}}^*) + 2{\rm Re} (M{ _{a^{3/2}{a_s}}}M_{a^{1/2}{a_s}}^*). \label{eq:1}
\eeq
Here, the first terms $d\G_1^0$, second two terms $d\G_2^{(0,1)}$ and the third two terms $d\G_3^{(0,1)}$ are the full QED contributions at LO level, the pure QCD contributions at NLO level and the cross terms up to NLO level , respectively. Meanwhile, the dimensional regularization $D = 4-2\epsilon$ is emploied to segregate the ultraviolet (UV) and infrared (IR) divergence. In the context of the on-mass-shell (OS) and $\rm \overline {MS}$ program, the renormalization constants $Z_m$, $Z_2$, $Z_g$ and $Z_3$ are used,
\beq
\dz{m}{OS} &=& - 3\CF\frac{\as}{4\pi}\bigg[\frac1{\epx{UV}} - \gE + \lnn{m_c} + \frac43\bigg],
\nn \\
\dz{2}{OS} & =& - \CF \frac{\as}{4\pi}\bigg[\frac1{\epx{UV}} + \frac{2}{\epx{IR}} - 3\gE + 3\lnn{m_c} + 4\bigg],
 \nn \\
\dz{3}{OS} &=&\frac{\as}{4\pi} \bigg[(\bo^{\prime} - 2\CA)\bigg(\frac1{\epsilon_{\rm UV}} + \frac1{\epx{IR}}\bigg) - \sum\limits_{Q=c,b}\frac43 \TF
\nn \\
&\times& \bigg(\frac1{\epx{UV}}- \gE + \lnn{m_Q}\bigg)\bigg],
\nn \\
\dz{g}{\overline {MS}} &=& - \frac{\bo}2 \frac{\as}{4\pi}\bigg[\frac1{\epx{UV}} - \gE + \ln (4\pi )\bigg],
\eeq
with $\gE$ denotes the Euler constant, $\bo = \frac{11\CA}{3} -\frac{4\TF n_f}{3}$ represents the one-loop $\beta$-function coefficient, and $\bo^\prime = \frac{11\CA}3 -\frac{4\TF n_{lf}}3$, where the $n_f$ ($n_{lf}$) stands for the active (light) quark flavor. In which $\TF=1/2$, $\CA=3$, $n_f=4$, and $n_{lf}=n_f-1=3$. \footnote{The expression can also be found in our previous work~\cite{Sun:2021hca}.}

In the detailed calculation, the toolchain we utilized are: {\tt FeynArts}~\cite{Hahn:2000kx} $\to$ {\tt FeynCalcFormLink}~\cite{Feng:2012tk} $\to$ {\tt Apart}~\cite{Feng:2012iq} $\to$ {\tt FIRE}\cite{Smirnov:2008iw} $\to$ {\tt X-package}. Precisely, we can obtain the Feynman diagrams and the corresponding hadron amplitude expressions with the FeynArts package; Further treatment and algebraic calculation of the Dirac and color matrices are performed by FeynCalcFormLink. The Apart function provides an additional simplification of the hadron amplitudes through partial fractions of the integrals for the IR divergence. The main integral can be readily obtained by using the FIRE package, which provides a complete IBP simplification of the hadron amplitudes by using a strategy based mainly on the Laporta algorith. Subsequently, we can be evaluated the main integral through using the X-package.
Noteworth, due to the special nature of $\g{5}$, we need to pay extra attention when the $D$-dimension $\gamma$ traces in $\hat \G_{\rm Loop}$ involve the $\g{5}$ matrix, and we utilize the following program~\cite{Berezhnoy:2021tqb, Hahn:1998yk} to handle it: For the existence triangle quark loops in NLO QCD correction diagrames, the expression of the quark loops always starts from Z-vertex and rewritten according to the definition of security of $\g{5}$ in the $D$-dimensional case, see Eq.~\eqref{eq:16},
\beq
\g{5}= - \frac{i}{24} \epsilon_{\alpha \beta \sigma \rho } \g{\alpha} \g{\beta} \g{\sigma} \g{\rho}, \label{eq:16}
\eeq
where $\epsilon_{\alpha \beta \sigma \rho }$ is 4 or $D$-dimensional Levi-Civita tensor. As a cross check, through choosing the same input parameters, we have revealed the NLO results of the process $\sigma(e^-e^+\to J/\psi+\eta_c(\chi_{cJ}))$ in Refs.~\cite{Sun:2018rgx,Jiang:2018wmv}, whose NLO QCD correction diagrams similar to the process $Z \to J/\psi+\eta_c(\chi_{cJ})$.

\btabs[t]
\caption{The total decay width for $Z\to J/\psi+H$ decays with $H = (\eta_c, J/\psi, \chi_{c0}, \chi_{c1}, \chi_{c2})$ (in unit: $10^{-12}~{\rm GeV}$), where the $\G_1^0$ stand for the full QED contributions at LO level, $\G_2^{0,1}$ represent the pure QCD contributions at NLO level and $\G_3^{0,1}$ are the cross terms up to NLO level, respectively. The total decay width are the sum of all sub-contributions, e.g. $\G_{\rm total}=\G_1^0+\G_2^0+\G_2^1+\G_3^0+\G_3^1$.}
\btabu{l l l l l l l l}
\hl \hl
Processes~~~~~~~~~~~&$\mu_r$~~~~~~~~~~~~~~&~~$\G_2^0$~~~~~~~~~~~~~~&~ $\G_1^0$~~~~~~~~~~~~~&~~~$\G_3^0$~~~~~~~~~~~~ &~~~$\G_2^1$~~~~~~~~~~~~~&~~~$\G_3^1$~~~~~~~~~~~~~~&$\G_{\rm total}$ \\ \hl
               & $2m_c$  & 0.798 & 33.86 &~10.40 &~2.990 &~18.19 & 66.24 \\
$J/\psi+\eta_c$    & $m_Z/2$ & 0.192 & 33.86 &~5.106 &~0.531 &~6.733 & 46.43 \\
               & $m_Z$   & 0.152 & 33.86 &~4.541 &~0.405 &~5.799 & 44.76 \\
\hl
               & $2m_c$  & 1.344 & 227.8 &~35.00 &~5.269 &~57.69 & 327.1 \\
$J/\psi+J/\psi$    & $m_Z/2$ & 0.324 & 227.8 &~17.19 &~0.922 &~21.81 & 268.1 \\ \
               & $m_Z$   & 0.256 & 227.8 &~15.29&~0.702 &~18.85 & 262.9 \\
\hl
               & $2m_c$ & 4.557 & 1.396 &~2.006 &~0.890 &~1.293 & 10.14\\
$J/\psi+\chi_{c0}$ & $m_Z/2$ & 1.099 & 1.396 &~0.982 &~1.115 &~0.763 & 5.355 \\
               & $m_Z$  & 0.870 & 1.396 &~0.874 &~0.966 &~0.695 & 4.799 \\ \hl
               & $2m_c$ & 0.197 & 8.364 & -0.009 &~0.287 &~2.094 & 10.93 \\
$J/\psi+\chi_{c1}$ & $m_Z/2$ & 0.047 & 8.364 & -0.004 &~0.078 &~0.503 & 8.988 \\ \
               & $m_Z$  & 0.038 & 8.364 & -0.004 &~0.062 &~0.398 & 8.858\\ \hl
               & $2m_c$ & 7.780 & 2.795 &~1.424 & -8.822 & -0.785 & 2.393\\
$J/\psi+\chi_{c2}$ & $m_Z/2$ & 1.877 & 2.795 &~0.700 &~0.679 &~0.132 & 6.182 \\
               & $m_Z$  & 1.485 & 2.795 &~0.622 &~0.787 &~0.169 & 5.858\\ \hl \hl
\etabu
\label{tabel:1}
\etabs

\section{Numerical results and discussions}\label{section:3}
The choices of the input parameters in doing the calculations are present as follows: The $m_c=1.5~{\rm GeV}$ and $m_Z=91.1876~{\rm GeV}$ indicates the mass of the charm quark and $Z$-boson, respectively. The choice of masses is $2m_c$ for $\eta_c$, $J/\psi$ and $\chi_{cJ}$ as to ensure the gauge invariance. We also adopt $|\Psi_{c\bar c}(0)| = \sqrt{1/4\pi}|R_S(0)|,$ and $|\Psi^\prime_{c\bar c}(0)| = \sqrt{3/4\pi}|R^\prime_P(0)|$, with $|R_S(0)|^2=0.81~{\rm GeV}^3$, $|R^\prime_P(0)|^2=0.075~{\rm GeV}^5$, which is consistent with Ref.~\cite{Eichten:1995ch}. The Weinberg angle $\theta_w$ is taken as $\theta_w={\rm arcsin}\sqrt {0.2312}$. The decay width of $Z$-boson is taken as $\G_Z=2.4952~\rm GeV$, which is identical to PDG~\cite{ParticleDataGroup:2018ovx}.

\btabs[ht]
\caption{The corresponding ratios ${\cal R}_i$ and ${\cal B}_{ri}$ with $i = (0,1,2)$ of each component for the processes $Z\to J/\psi+H$ decays with $H = (\eta_c, J/\psi, \chi_{c0}, \chi_{c1}, \chi_{c2})$.}
\btabu{l l l l l l l l l}
\hl \hl
Processes~~~~~~~~~& $\mu_r$~~~~~~~~~~&~\,${\cal R}_{0}$~~~~~~~~~&~\,${\cal R}_{1}$~~~~~~~~&~~\,${\cal R}_{2}$~~~~~~~~&~~~~~~${\cal B}_{r0}$~~~~~~~~~~~~~&~~~~~~${\cal B}_{r1}$~~~~~~~~~~~~~~~&~~~~~~${\cal B}_{r2}$ \\
\hl
              & $2m_c$  & 0.470 & 0.859 & ~2.745 & $3.198\times 10^{-13}$ & ~$1.518\times 10^{-12}$ & $2.655\times 10^{-11}$ \\
$J/\psi+\eta_c$   & $m_Z/2$ & 0.185 & 0.927 & ~7.058 & $7.712\times 10^{-14}$ & ~$2.899\times10^{-13}$ & $1.861\times10^{-11}$ \\
              & $m_Z$  & 0.161 & 0.935 & ~8.145 & $6.102\times 10^{-14}$ & ~$2.235\times 10^{-13}$ & $1.794\times 10^{-11}$ \\
\hl
              & $2m_c$  & 0.238 & 0.916 & ~5.292 & $5.386\times10^{-13}$ & ~$2.650\times10^{-12}$ & $1.311\times10^{-10}$ \\
$J/\psi+J/\psi$    & $m_Z/2$ & 0.093 & 0.959 & ~13.79 & $1.299\times10^{-13}$ & ~$4.990\times10^{-13}$ & $1.074\times10^{-10}$ \\
              & $m_Z$  & 0.080 & 0.964 & ~15.95 & $1.028\times10^{-13}$ & ~$3.842\times10^{-13}$ & $1.054\times10^{-10}$ \\
\hl

              & $2m_c$  & 0.274 & 0.592 & ~0.367 & $1.826\times10^{-12}$ & ~$2.183\times10^{-12}$ & $4.062\times10^{-12}$\\
$J/\psi+\chi_{c0}$ & $m_Z/2$ & 0.540 & 0.406 & ~0.444 & $8.874\times10^{-13}$ & ~$2.146\times10^{-13}$ & $1.054\times10^{-12}$ \\
              & $m_Z$   & 0.529 & 0.418 & ~0.476 & $3.485\times10^{-13}$ & ~$7.355\times10^{-13}$ & $1.923\times10^{-12}$ \\ \hl
              & $2m_c$  & 0.278 & 0.879 & -0.018 & $7.878\times10^{-14}$ & ~$1.939\times10^{-13}$ & $4.382\times10^{-12}$ \\
$J/\psi+\chi_{c1}$ & $m_Z/2$ & 0.069 & 0.866 & -0.035 & $1.900\times10^{-14}$ & ~$5.009\times10^{-14}$ & $3.602\times10^{-12}$ \\ \
              & $m_Z$   & 0.055 & 0.864 & -0.039 & $1.503\times10^{-14}$ & ~$4.004\times10^{-14}$ & $3.550\times10^{-12}$\\ \hl
              & $2m_c$  & -0.801 & 0.082 & -1.367 & $3.118\times10^{-12}$ & -$4.175\times10^{-13}$ & $9.589\times10^{-13}$\\
$J/\psi+\chi_{c2}$ & $m_Z/2$ & 0.151 & 0.163 & ~0.274 & $7.521\times10^{-13}$ & ~$1.024\times10^{-12}$ & $2.478\times10^{-12}$ \\
              & $m_Z$   & 0.195 & 0.177 & ~0.274 & $5.950\times10^{-13}$ & ~$9.104\times10^{-13}$ & $2.347\times10^{-12}$\\ \hl \hl
\etabu
\label{tabel:3}
\etabs
According to the parameters mentioned before, the total decay width for exclusive doubly charmonium production via $Z^0$-boson decay is shown in Table~\ref{tabel:1}. One can readily obtain the following:
\begin{itemize}
\item After introducing the complete QED diagrams, the total decay widths of the production of $ J/\psi+J/\psi$ pair and $J/\psi+\eta_c$ pair via $Z$-boson decay are in the order of $10^{-10}$ and $10^{-11}$ at the LO level, respectively, which is comparable with the results given in Ref.~\cite{Gao:2022mwa}. Moreover, the contribution of the cross term is one order of magnitude larger than that of pure QCD at LO level for process $Z\to J/\psi+J/\psi(\eta_c)$, both at LO and NLO level, which indicates the non-negligibility and necessity of considering the interference effect.
\item Compared to $J/\psi+J/\psi(\eta_c)$, the case of $J/\psi +\chi_{c0}(\chi_{c2})$ becomes more moderate. The contributions from the full QED process, cross terms and QCD process are in the $10^{-12}$ order at Tree-level. Precisely, the contributions from the QED and cross terms can reach $36\%$($30\%$) and $11\%$($43\%$) of pure QCD process, respectively. Meanwhile, the contribution of the newly considering contribution of interference term at NLO level, namely $\G_3^1$ can also reach $28\%$ of pure QCD at LO level.
\item In the case of $\chi_{c1}$, the contribution of the complete QED is one order of magnitude larger than that of QCD at Tree-level. Moreover, the decay width of the higher order terms for the cross term, namely $\G_3^1$ can up to $10^{-11}$ order, and the pure QCD process in $10^{-12}$ order at NLO level, which demonstrates the significance of considering the NLO QCD corrections to interference term.
\end{itemize}

\bfigs[t]
\centering
\grap{0.32}{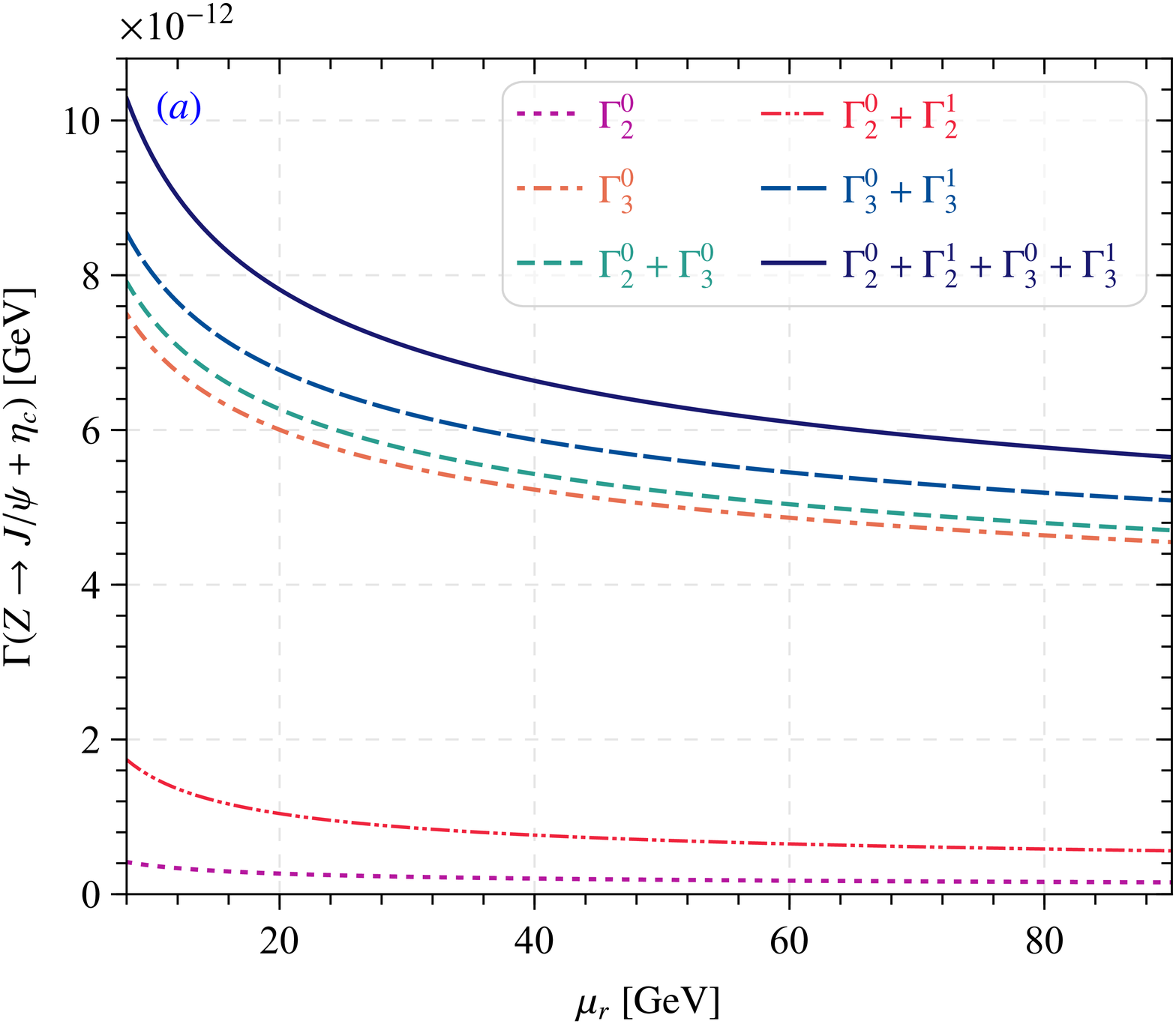}~\grap{0.32}{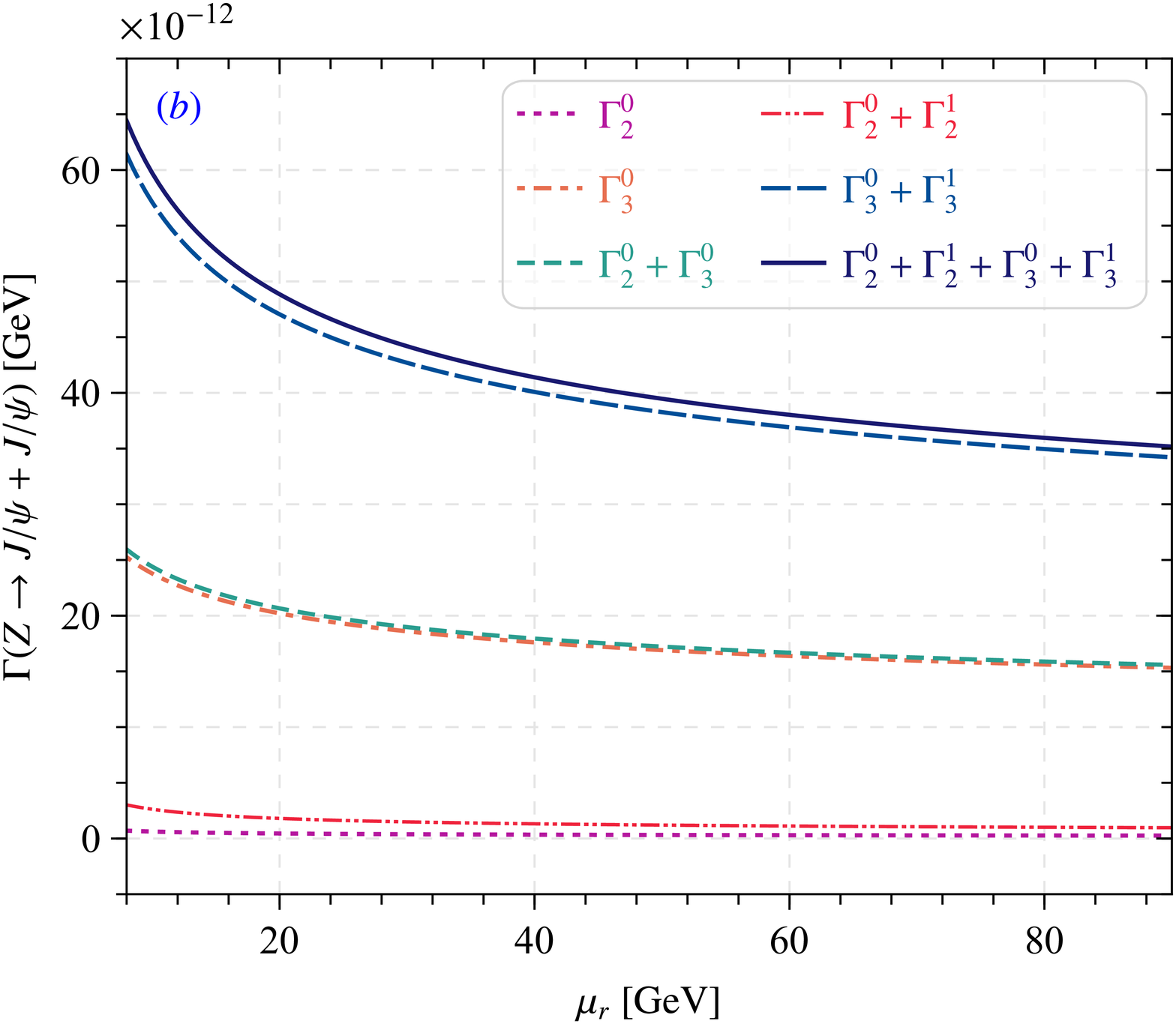}~\grap{0.32}{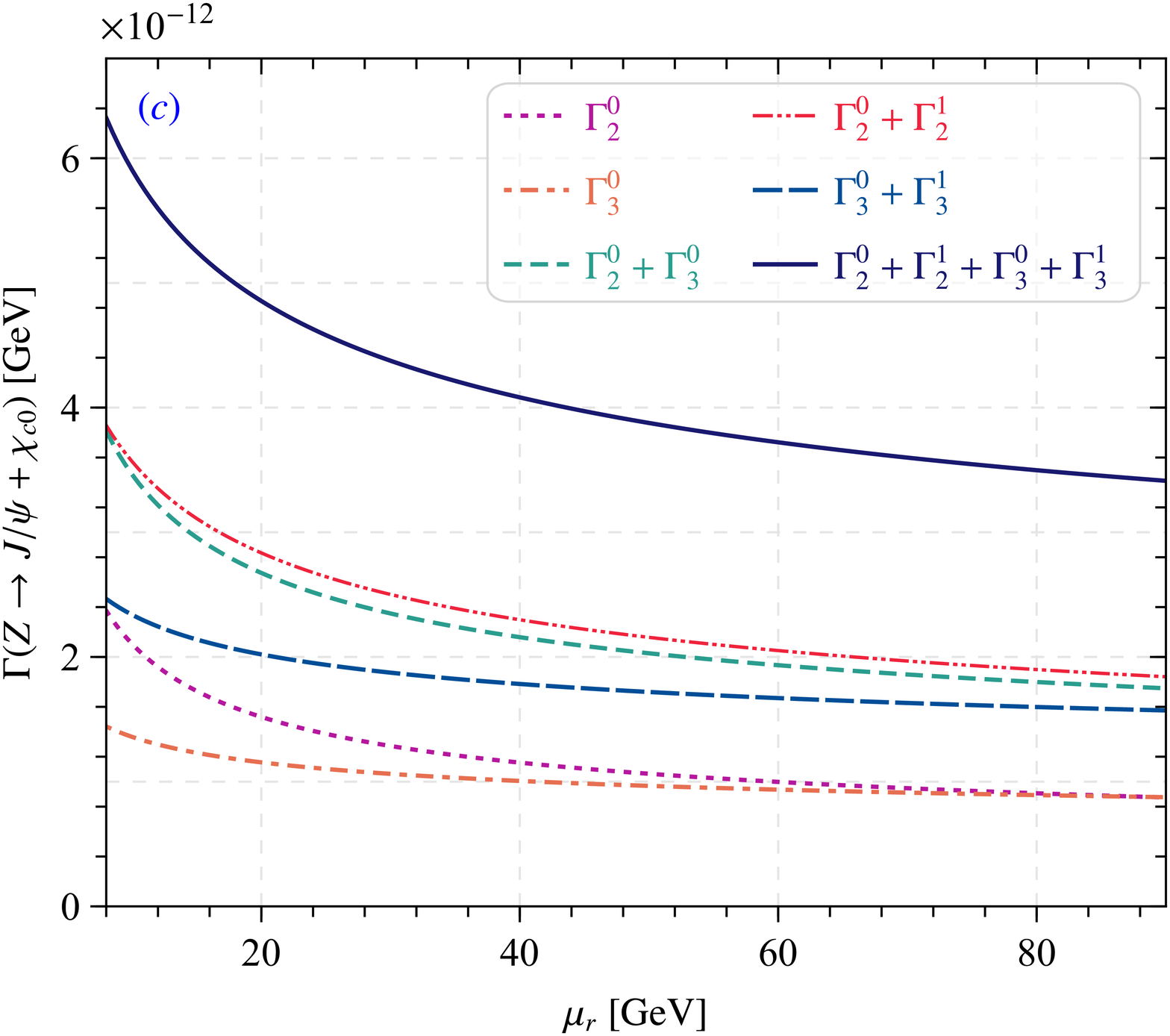}\\
\grap{0.32}{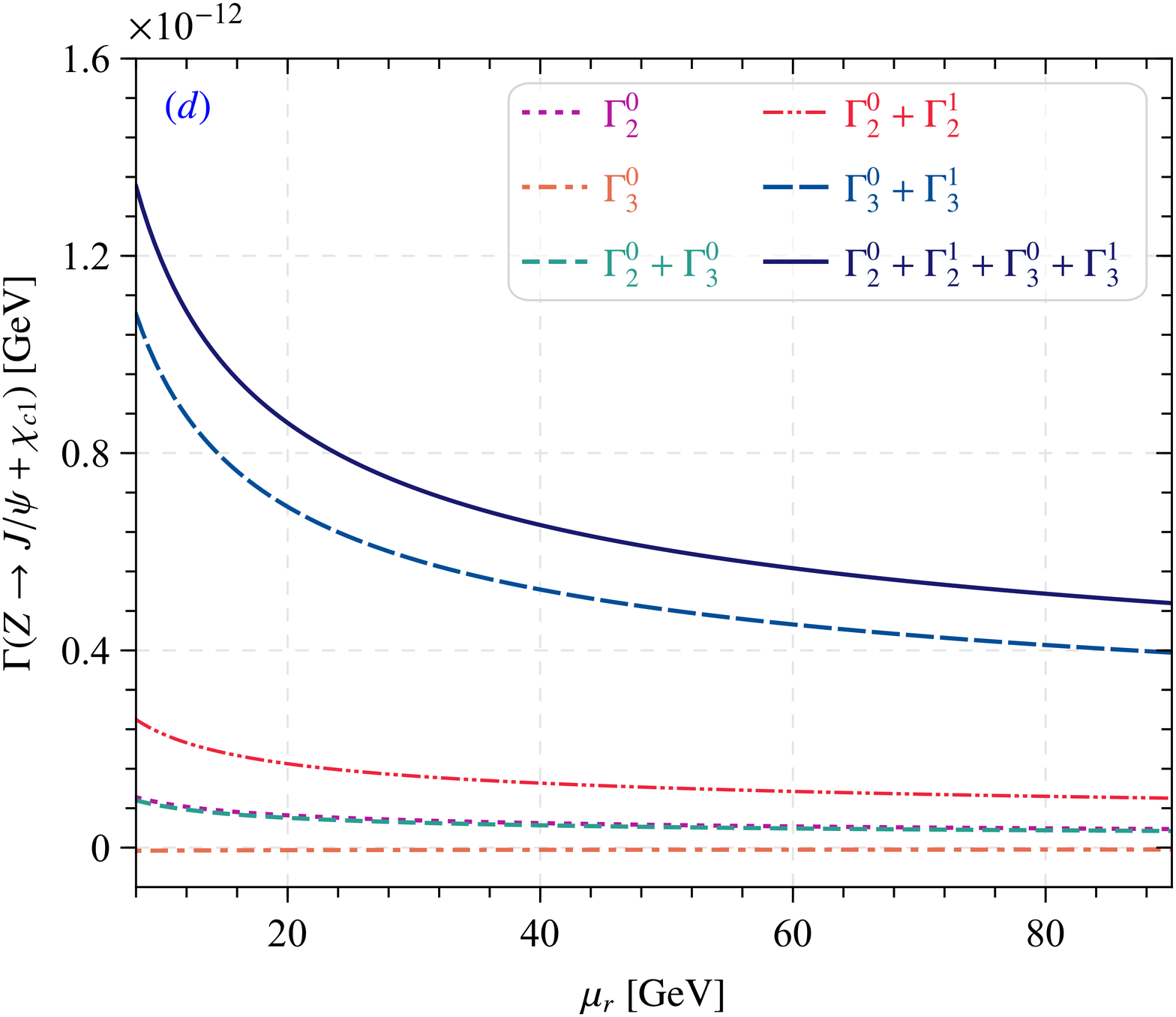}~\grap{0.32}{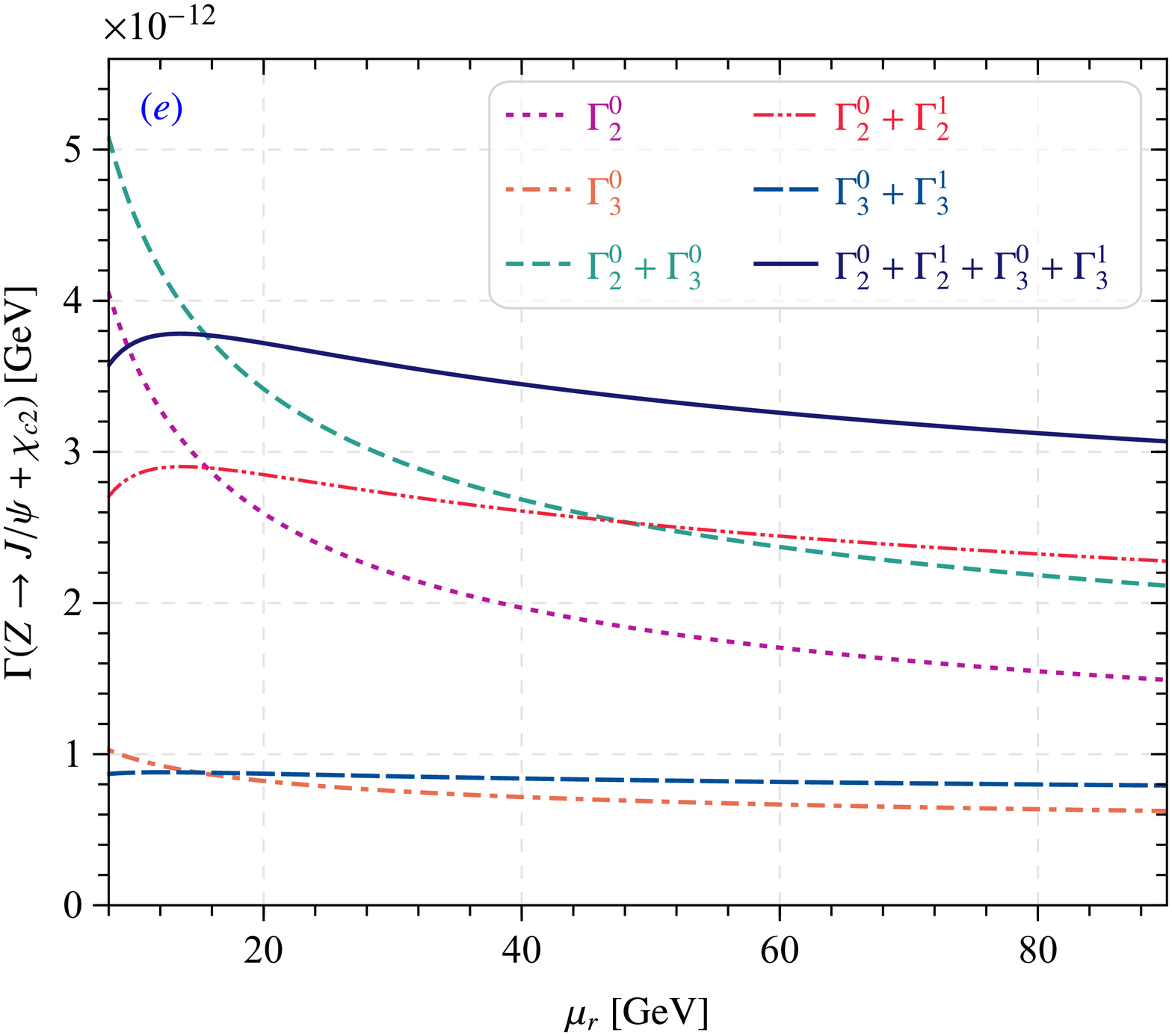}
\caption{(Color online) The six different decay widths for the $Z\to J/\psi+H$ processes with $H = (\eta_c, J/\psi, \chi_{c0}, \chi_{c1}, \chi_{c2})$. In which the charmed quark mass is taken as $m_c=1.5~{\rm GeV}$.}\label{fj4}
\efigs

To further investigate effects of each component, one can usually take the following ratios, which can be read off
\beq
&& {\cal R}_0 =(\G_2^1+\G_3^1)/(\G_1^0+\G_2^0+\G_3^0),
\nn\\
&& {\cal R}_1 = \G_3^1/(\G_2^1+\G_3^1),
\nn\\
&& {\cal R}_2 = \G_3^0/(\G_2^0+\G_2^1),
\nn\\
&& {\cal B}_{r0} = \G_2^0/\G_Z,
\nn\\
&& {\cal B}_{r1} = (\G_2^0+\G_2^1)/\G_Z,
\nn\\
&& {\cal B}_{r2} = (\G_2^0+\G_2^1+\G_3^0+\G_3^1+\G_1^0)/(\G_Z),
\eeq
respectively. We present the ratios ${\cal R}_i$ and ${\cal B}_{ri}$ with $i = (0,1,2)$ in Table~\ref{tabel:3}, which can be seen that:
\begin{itemize}
\item Up to NLO level, the higher order terms $\G_2^1+\G_3^1$ can effect the evaluations for LO. Especially, for the cases of $J/\psi+\eta_c(\chi_{c0})$, which can even reach $47\%(53\%)$, the corresponding renormalization scale is $2m_c(m_Z)$. Thus, the contributions from the NLO level should be taken into consideration in our calculations.
\item In most instances the one-loop QCD corrections for the cross terms plays an essential role in higher order terms, except for the case of $J/\psi+\chi_{c2}$, which is only about $10\%$. The ratio ${\cal R}_{1}$ in the Table~\ref{tabel:3}, which demonstrate the significance and necessity of considering the NLO correction for the interference term.
\item After introducing the complete QED diagrams, the branching ratio ${\cal R}_{2}$ of the $ J/\psi+ J/\psi(J/\psi+\eta_c)$ pair via $Z$-boson decay could up to $10^{-10}(10^{-11})$ at the NLO level. While, its only about $10^{-12}(10^{-13})$ order for pure QCD contributions. The tension between ${\cal B}_{r1}$ and ${\cal B}_{r2}$ indicates the significance of considering the NLO correction for the cross terms effect between the QCD and complete QED diagrams.
\end{itemize}
To analysis the behaviors of the $Z\to J/\psi+J/\psi(\eta_c,\chi_{cJ})$ decay width running with the renormalization scale $\mu_r$, we present the curves in Fig.~\ref{fj4}, which indicate that the numerous gap between the pure QED process and the pure QCD process at the LO level, the QED process that has been overlooked can have a significant impact on our prediction results, especially for $J/\psi+\eta_c(J/\psi)$. Moreover, there is also exist a huge gap between the pure QCD process and interference terms at NLO level for $J/\psi+J/\psi$, and the similar situation exists in $J/\psi+\eta_c(\chi_{c1})$,  Thus, the contributions from the cross terms between the QED and QCD diagram should be taken into consideration in our calculations, at NLO level. Meanwhile, the $\G_3^{0}+\G_3^1$ will significantly affect the results of the width decay for the $J/\psi+\chi_{c0}(\chi_{c2})$. Furthermore, the gap between $\G^0_2(\G^0_3)$ and $\G^0_2+\G^1_2(\G^0_3+\G^1_3)$ indicates that NLO QCD correction is necessary.

\bfig[t]
\centering
\grap{0.42}{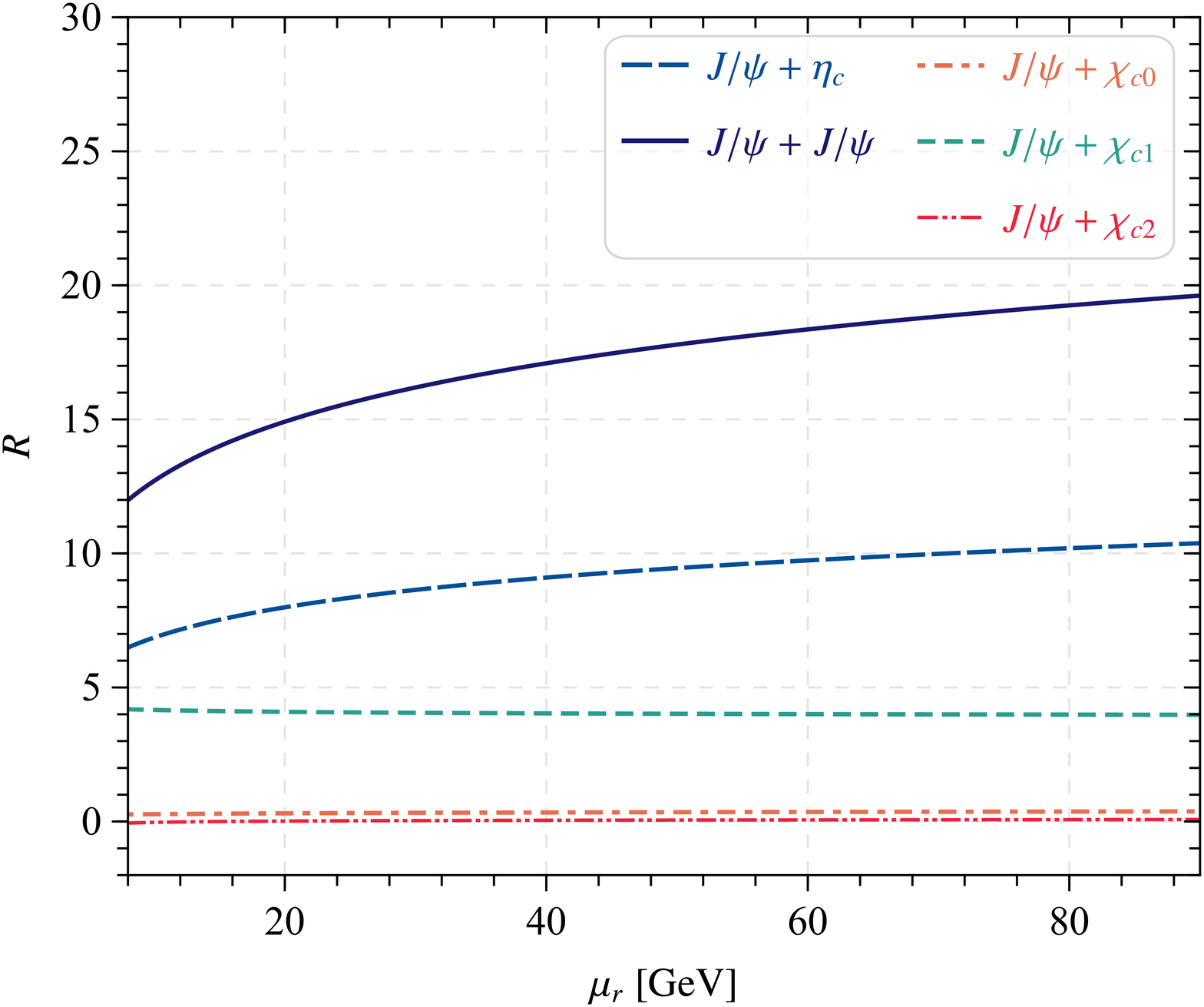}
\caption{(Color online) The renormalization scale $\mu_r$ as a function of $R$, where $R=\G_3^1/(\G_2^0 + \G_2^1)$ with $\G_2^{0,1}$ denote the contributions of pure QCD process at NLO level.}\label{fj5}
\efig
Furthermore, in order to investigate the relative necessity of the newly calculated cross term at NLO level, one can define the ratio
\beq
R = \G_3^1/(\G_2^0+\G_2^1).
\label{eq:Ratio}
\eeq
Then, we present the $R$ changed with the reorganization scale $\mu_r$ in Fig.~\ref{fj5}. As the figure demonstrated, the element of $\G_3^1$ will play an increasingly important role as the renormalization scale $\mu_r$ rises, especially for the $J/\psi+J/\psi (\eta_c,\chi_{c1})$ case. Specifically, when $\mu_r = m_Z$, the predicts of $R$ for the indirect production of $J/\psi+J/\psi$, $J/\psi+\eta_c$ and $J/\psi+\chi_{c1}$ pair via $Z$-boson can reach up to $19$, $10$ and $4$, respectively. Precisely, for the indirect production of doubly charmonium via $Z$-boson decay, the one-loop QCD corrections for the cross terms influence may be essential, or even crucial in comparison with the pure QCD process at NLO level.\\

\section{Summery}\label{section:4}
To alleviate the tension between theoretical results and CMS collaboration data ${\cal B}_{Z\to J/\psi J/\psi}$, we performed a further research on the process $ Z\to J/\psi+J/\psi$ by introducing the complete QED diagrams. It is revealed that in addition to the results of cross term between QED and QCD diagram at LO level, NLO QCD correction of the interference term can also enormously elevate its LO results, which demonstrate the necessity and significance of considering the NLO QCD correction of the interference term. With these all contributions in mind, the gap between theory predicts and CMS Collaboration data remain significant. Meanwhile, in order to provide a forward-looking guide to the experiment, we also investigate the process $ Z\to J/\psi+\eta_c(\chi_{cJ})$ at the NLO level. For the indirect production of the $J/\psi+\eta_c$, the situation is similar to $ Z\to J/\psi+J/\psi$, the cross term between QED and QCD diagram, and NLO QCD correction of the interference term can also greatly enhance the results of LO level. In the case of $J/\psi+\chi_{cJ}$, NLO QCD correction of the interference term can largely affect the total width decay at LO level. Thus, in order to achieve a reasonable estimate of the total decay width of $Z\to J/\psi+J/\psi(\eta_c,\chi_{cJ})$ , it is indispensable to consider the contribution of NLO QCD correction of the interference term.

\acknowledgments
We are grateful for the Professor Zhan Sun's valuable comments and suggestions. This work is supported in part by the Natural Science Foundation of China under Grant No. 12265010, and by the Project of Guizhou Provincial Department of Education under Grant No. KY[2021]030.

\end{document}